\pdfoutput=1
\documentclass[prd,twocolumn,aps,superscriptaddress,showpacs,preprintnumbers,nofootinbib]{revtex4-1}
\usepackage{latexsym,amsmath,amssymb}
\usepackage{graphicx}
\usepackage[table,usenames,dvipsnames]{xcolor}
\usepackage{hyperref}
\definecolor{azure(colorwheel)}{rgb}{0.0, 0.5, 1.0}
\hypersetup{colorlinks=true,linkcolor=blue,citecolor=azure(colorwheel),pdfborder={0 0 0}}
\usepackage[utf8]{inputenc}
\usepackage{feynmp}
\usepackage{bbold}
\usepackage{fontawesome}
\newcolumntype{P}[1]{>{\centering\arraybackslash}p{#1}}
\usepackage[normalem]{ulem}

\makeatletter
\newcommand*{\rom}[1]{\expandafter\@slowromancap\romannumeral #1@}
\makeatother

\newcommand{\ignore}[1]{}

\begin{document}
\title{
Axion Magnetic Resonance:\\
A Novel Enhancement in Axion-Photon Conversion
}
\author{Hyeonseok Seong}\email{hyeonseok.seong@desy.de}
\affiliation{Deutsches Elektronen-Synchrotron DESY, Notkestr. 85, 22607 Hamburg, Germany}

\author{Chen Sun}\email{chensun@lanl.gov}
\affiliation{Theoretical Division, Los Alamos National Laboratory, Los Alamos, NM 87545, USA}

\author{Seokhoon Yun}\email{seokhoon.yun@ibs.re.kr}
\affiliation{Particle Theory  and Cosmology Group, Center for Theoretical Physics of the Universe, Institute for Basic Science (IBS), Daejeon, 34126, Korea}
\affiliation{Dipartimento di Fisica e Astronomia, Università degli Studi di Padova, Via Marzolo 8, 35131
  Padova, Italy}
\affiliation{INFN, Sezione di Padova, Via Marzolo 8, 35131 Padova, Italy}

\date{\today}

\begin{abstract}
We identify a new resonance, axion magnetic resonance (AMR), that can greatly enhance the conversion rate between axions and photons.  A series of axion search experiments rely on converting them into photons inside a constant magnetic field background. A common bottleneck of such experiments is the conversion amplitude being suppressed by the axion mass when $m_a \gtrsim 10^{-4}~$eV.  We point out that a spatial or temporal variation in the magnetic field can cancel the difference between the photon dispersion relation and that of the axion, hence greatly enhancing the conversion probability. 
We demonstrate that the enhancement can be achieved by both a helical magnetic field profile and a harmonic oscillation of the magnitude. Our approach can extend the projected ALPS II reach in the axion-photon coupling ($g_{a\gamma}$) by two orders of magnitude at $m_a = 10^{-3}\;\mathrm{eV}$ with moderate assumptions.~\href{https://github.com/ChenSun-Phys/axion-magnetic-resonance}{\faGithub}
\end{abstract}


\preprint{DESY-23-118}
\preprint{LA-UR-23-29378}
\preprint{CTPU-PTC-23-39}

\maketitle



\textbf{Introduction}~~ While there are a plethora of experimental agendas to search for axions utilizing {its coupling to photon} {independent of their cosmological relic abundance~\cite{CAST:2007jps,CAST:2017uph,Ehret:2010mh,Bahre:2013ywa,Ortiz:2020tgs,OSQAR:2015qdv,DellaValle:2015xxa,Betz:2013dza} or leveraging the relic abundance by assuming axions to be dark matter \cite{ADMX:2021nhda,Alesini:2020vny,Alesini:2019ajt,Alesini:2022lnp,Crisosto:2019fcj,DePanfilis:1987dk,Devlin:2021fpqa,Gramolin:2020ict,Gramolin:2020ictSalemi:2021gck,McAllister:2017lkb,Quiskamp:2022pks,Yang:2023yry}} {(see Refs.~\cite{Agrawal:2017cmd,Irastorza:2018dyq,Choi:2020rgn} for comprehensive reviews)}, most bounds get considerably weakened at $m_a \gtrsim 10^{-4}-10^{-2} ~\mathrm{eV}$. In this mass range, experiments that rely on an axion-to-photon conversion in a magnetic field battle the so-called {small-mixing-non-linear} regime, where the conversion probability is suppressed by {$m_a^{-4}$} much before it is limited by the length scale of the experiments. Therefore, in this regime, a longer conversion baseline will not improve the bound. 
It is of great interest to study ways of enhancing the axion-photon conversion in this regime.

The usual axion-photon oscillation formula assumes a coherent magnetic field that is constant in both the amplitude and its orientation~\cite{Sikivie:1983ip,Anselm:1985obz,VanBibber:1987rq}. 
{
The effect of a non-trivial magnetic profile is only briefly mentioned in the seminal works~\cite{Sikivie:1983ip,VanBibber:1987rq}.
In this Letter, we perform a critical study on spatial or temporal variation of the magnetic field. We show that the magnet variation can compensate the fast axion-photon oscillation due to a large axion mass, leading to a resonantly enhanced axion-photon conversion. 
In particular, for the first time, we identify the optimal spatial/temporal magnetic profile that leads to the maximal enhancement.
We then apply this resonance to Light-Shining-Through-Walls (LSTW) experiments, model-independent axion searches that do not rely on an axion dark matter abundance.
}

The spatial variation of the magnetic field necessitates a more comprehensive treatment of the dynamics of the system.
One way to understand this resonance is to use the language of parametric resonance. It occurs when the frequency of the spatial variation coincides with the axion-photon oscillation frequency. This resonance phenomenon holds the potential to significantly amplify the conversion probability, providing an exciting avenue for experimental exploration.

Alternatively, the enhancement can be understood if we change to the basis where the magnetic field becomes constant. An extra mass-like term for the photon states is generated during this {transformation}. The enhancement is a result of an avoided level crossing.~\footnote{Similar effect {due to a varying matter potential in the neutrino Mikheyev–Smirnov–Wolfenstein  effect is well studied in} \textit{e.g.}~\cite{Kuo:1989qe}.}
This contribution vanishes for a constant mixing matrix, hence, it has been safely neglected in the well-established oscillation formula (e.g.~\cite{Raffelt:1987im,Mirizzi:2005ng,Mirizzi:2006zy,Mirizzi:2009nq, Masaki:2017aea,Buen-Abad:2020zbd}).

Lastly, we provide a third approach by noting that there is a mismatch between the dispersion relations of axions and photons. The resonance happens when the magnetic field variation frequency coincides with this gap, {i.e., the momentum transfer}. For this reason, we dub the new resonance axion magnetic resonance (AMR), by analogy with the nuclear magnetic resonance (NMR). 

As a demonstration of the experimental relevance of the AMR, we take the LSTW experiment ALPS II~\cite{Bahre:2013ywa,Ortiz:2020tgs} as an example that a helical magnetic profile can enhance the {experimental} reach by $1$-$2$ orders of magnitude where the search remains model-independent and does not require a local axion abundance.

\textbf{Helical Magnetic Background~~}
One of the most intriguing axion interactions in terms of phenomenological searches is its anomalous coupling with the photon
\begin{align}
  \mathcal{L}_{a\gamma}
  & =
     \frac{g_{a\gamma}}{4}a F_{\mu\nu} \tilde F^{\mu\nu} = - g_{a\gamma}a\,\mathbf{E}\cdot \mathbf{B},
\end{align}
where $\tilde F^{\mu\nu} = \epsilon^{\mu\nu\alpha\beta} F_{\alpha\beta}/2$.
We denote the propagation direction as $z$, and the photon-axion system in the \textit{interaction} basis to be $ \tilde \Psi = [\gamma_\perp ~  \gamma_\parallel ~  a]^T$. The $\perp(\parallel)$ is for the direction of photon polarization perpendicular (parallel) to the magnetic field at {the initial point}, $z=0$, at rest with the lab. 

Let us start with $\mathbf{B}$ of a constant magnitude and a changing orientation along the $z$-direction, \textit{i.e.}, a helical profile. Since the $\perp$ and $\parallel$ directions are defined in terms of the $\mathbf{B}$ field direction at $z=0$, when $\mathbf{B}$ changes direction it will have both $\perp$ and $\parallel$ components, \textit{i.e.} $\mathbf{B} = [B_\perp, B_\parallel, 0] = [B\sin\theta, B\cos\theta, 0]$, with $\theta(z=0) = 0$. 
The equation of motion (EOM) reads
\begin{align}
  \label{eq:rotating-B-int-basis}
  &   i \partial_z
    \tilde \Psi
   =
    \frac{1}{2\omega} H^2(\theta)
    \tilde \Psi   
    \\
  & =
    \frac{1}{2\omega}
    \begin{bmatrix}
      0 & 0 & g_{a\gamma}  \omega B \,{\rm s}_\theta \\
      0 & 0 & g_{a\gamma} \omega B \,{\rm c}_\theta\\
      g_{a\gamma} \omega B \,{\rm s}_\theta & g_{a\gamma}  \omega B \,{\rm c}_\theta  & \Delta m^2_{a\gamma}
    \end{bmatrix}
    \begin{bmatrix}
      \gamma_\perp \\
      \gamma_\parallel\\
      a
    \end{bmatrix},
    \notag
\end{align}
where ${\rm c}_\theta \equiv \cos\theta$, ${\rm s}_\theta\equiv \sin\theta$, and $\Delta m^2_{a\gamma} \equiv m_a^2 - m_\gamma^2$. We have subtracted the diagonal term $m_\gamma^2$ since it only generates an overall phase for $\tilde \Psi$. {
We first make a rotation in the {$\gamma_\perp$-$\gamma_\parallel$} direction:
\begin{align}
  & H^2(\theta)
   =
    U(\theta) H^2(0) U^\dagger(\theta)
    \\
  & = 
    \begin{bmatrix}
      {\rm c}_\theta & {\rm s}_\theta & 0 \\
      - {\rm s}_\theta & {\rm c}_\theta & 0 \\
      0 & 0 & 1
    \end{bmatrix}
    \begin{bmatrix}
      0 & 0 & 0 \\
      0 & 0 & g_{a\gamma} \omega B\\
      0 & g_{a\gamma}  \omega B   & \Delta m^2_{a\gamma}
    \end{bmatrix}    
    \begin{bmatrix}
      {\rm c}_\theta & -{\rm s}_\theta & 0 \\
      {\rm s}_\theta & {\rm c}_\theta & 0 \\
      0 & 0 & 1
    \end{bmatrix}.
    \notag
\end{align}
We define an auxiliary basis, $\hat \Psi = U^\dagger \tilde \Psi$. 
Since the rotation matrix $U$ is $z$-dependent, it can only diagonalize the Hamiltonian instantaneously. As a result, at different locations, we need different $U(\theta)$ matrices to transform $\tilde \Psi$ to $\hat \Psi$. This generates extra off-diagonal terms~\cite{Kuo:1989qe,Wang:2015dil} similar to a gauge transformation, while a $\theta(z)$ profile corresponds to a specific gauge fixing imposed by the external magnetic field:
\begin{align}
\label{eq:U-rot}
  i \partial _z \hat \Psi
  & = \bigg [
    U^\dagger \frac{H^2}{2\omega} U
    -
    i U^\dagger \partial _z U
    \bigg ] \hat \Psi
  \cr
  & =
    \begin{bmatrix}
      0 & -i \dot \theta & 0 \\
      i \dot \theta & 0 & g_{a\gamma}   B/2 \\
      0 & g_{a\gamma}  B/2  & \Delta m_{a\gamma}^2/2\omega
    \end{bmatrix}
    \hat \Psi\,,
\end{align}
where {$\dot \theta \equiv d \theta/dz$}. To diagonalize the 1-2 component, we perform a $45^\circ$ rotation in the 1-2 direction followed by a $90^{\circ}$ phase {shift} for the first component:
\begin{align}
  \label{eq:diagonalize-12}
  V =
  \begin{bmatrix}
    i & 0 & 0 \\
    0 & 1 & 0 \\
    0 & 0 & 1
  \end{bmatrix}
  \begin{bmatrix}
    \cos (\pi/4) & -\sin (\pi/4) & 0 \\
     \sin (\pi/4) & \cos (\pi/4) & 0 \\
    0 & 0 & 1
  \end{bmatrix},
\end{align}
which gives us the equation of motion for $\Psi = V^\dagger U^\dagger \tilde \Psi$
\begin{align}
  \label{eq:explicit-show-resonance-helical}
  i \partial _z
  \Psi
  & = 
    \begin{bmatrix}
      - \dot \theta & 0 & g_{a\gamma}  B/2\sqrt 2 \\
      0 & \dot \theta &  g_{a\gamma}  B/2\sqrt 2 \\
        g_{a\gamma}  B/2\sqrt 2 & g_{a\gamma}  B/2\sqrt 2  & \Delta m_{a\gamma}^2/2\omega 
    \end{bmatrix}
    \Psi \, .
\end{align}
In this basis,  we observe that a resonance can happen when $|\dot \theta | =  |\Delta m_{a\gamma}^2|/2\omega$, which leads to an amplified conversion probability. We denote this resonance as the axion magnetic resonance (AMR).

In the presence of a constant $\dot \theta$, and under the limit of $m_a^2/2\omega \gg g_{a\gamma} B$, the $a$-to-$\gamma$ conversion probability is given by~\footnote{This is an approximation in the small-mixing limit, so the unitarity is preserved up to the order of $g_{a\gamma}B/(\Delta m_{a\gamma}^2/2\omega)$ level.}
\begin{align}
\label{eq:conversion-prob-varyingB}
P_{a{\rightarrow} \gamma}
& \simeq 
\sum_{i=\pm}\frac{(g_{a\gamma}B/\sqrt{2})^2}{\Delta_i^2}
  \sin^2 \left (\frac{\Delta_i l}{2} \right) \,
\end{align}
with
\begin{align}
\label{eq:Delta-p-m}
\Delta_\pm = \sqrt{\left(\Delta m_{a\gamma}^2/2\omega \pm \dot{\theta}\right)^2 + \left(g_{a\gamma} B/\sqrt{2}\right)^2} \, .
\end{align}
{This is the well-known formula for a two-level quantum system going through the Rabi cycles, which is also used to model both NMR and qubits. }
Interestingly, $\pm\dot{\theta}$ can compensate the $\Delta m_{a\gamma}^2/2\omega$ term. Such a cancellation can enhance the oscillation amplitude or bring the conversion back to the linear regime{,  \textit{i.e.}, $\Delta_i l \ll 1$}. In both cases, it lifts the $m_a^{-4}$ mass suppression.
In Fig.~\ref{fig:num-vs-ana}, we show the numerical analysis of Eq.~\eqref{eq:rotating-B-int-basis} with a rotating magnetic field
and compare it with the conventional constant magnetic field setup (\textit{i.e.}, $\dot\theta=0$). We find good agreement between the numerical results and Eq.~\eqref{eq:conversion-prob-varyingB}.
In the limit $\dot\theta\rightarrow 0$, Eq.~\eqref{eq:conversion-prob-varyingB} approximately reduces to the usual oscillation formula; see Appendix~\ref{sec:axion-photon-usual-formula} for a brief review.
While there is a small difference in the wavenumber, it is relevant only in the maximal-mixing non-linear regime, which the LSTW experiments never enter. 
\begin{figure}[t!]
  \centering
  \includegraphics[width=.49\textwidth]{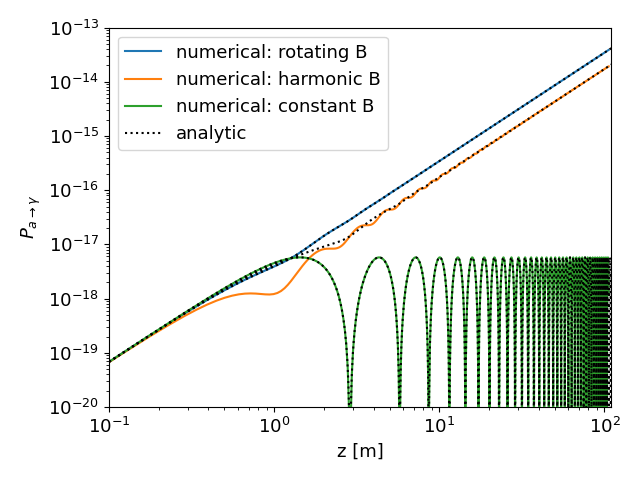}
  \caption{The axion-to-photon conversion probability with the following benchmark: $m_a = 10^{-3}\,\mathrm{eV}$, $\omega=1.16\,\mathrm{eV}~(1064\,\mathrm{nm}$), $g_{a\gamma} = 10^{-9}\,\mathrm{GeV}^{-1}$, $B = 5.3\,\mathrm{T}$. 
  The magnetic field is chosen to be constant (green), rotating as in Eq.~\eqref{eq:rotating-B-int-basis} with angular velocity $\dot \theta = m_a^2/2\omega$ (blue), and harmonic as in Eq.~\eqref{eq:harmonic-magnetic-field} with the same angular velocity $\dot \theta = m_a^2/2\omega$ (orange). 
  The solid curves are solved numerically, while the dotted curves are from the analytic expression Eq.~\eqref{eq:conversion-prob-varyingB} with the wave number modified according to Eq.~\eqref{eq:Delta-p-m}. 
  }
  \label{fig:num-vs-ana}
\end{figure}

The $\gamma$-to-$a$ conversion is more subtle {because it depends on the initial polarization of the photon.  
}
Let us look at the $a\rightarrow \gamma$ process instead and utilize the CPT theorem.
In the case of $\dot\theta = \Delta m_{a\gamma}^2/2\omega$, ${\Psi=}  [0~1~0]^{\rm T}$ will be the dominant photon component produced in the non-linear regime.
After rotating back with $U$ and $V$, it corresponds to $[-i/\sqrt 2 ~ 1/\sqrt 2 ~ 0]^{\rm T}$ in the $[\gamma_\perp ~ \gamma_\parallel~ a]^{\rm T}$ lab frame. 
This is expected since the axion field has no preferred direction. Heuristically, the external magnetic field determines the polarization of the daughter photons converted from axions. Since only the photons that are parallel to $\mathbf{B}$ couple to axions, with a rotating $\mathbf{B}$, the polarization vector of the signal photons corotates with $\mathbf{B}$.
This feature could potentially open up new search strategies. We leave a study of this to future work.

Without modifying the linearly polarized laser setup of LSTW, the axion production rate $P_{\gamma{\rightarrow} a} \simeq  P_{a {\rightarrow} \gamma}/2$ holds in the non-linear regime, regardless of the initial photon polarization; see Appendix~\ref{sec:appendix-ic} for more details. 
We report that $P_{\gamma{\rightarrow} a} < P_{a{\rightarrow} \gamma}$ in the linear regime unless we start with the initial polarization ${\Psi=}[0~1~0]^{\rm T}$,  {in which case $P_{\gamma{\rightarrow} a} = P_{a{\rightarrow} \gamma}$.}
The dependence on the initial photon polarization does not affect our main result since the major improvement for ALPS II is in the non-linear regime. 

 As a cross check of the formalism, we provide a derivation using the approach of parametric resonance in Appendix~\ref{sec:parametric} and another heuristic approach by changing of reference frame in Appendix~\ref{sec:alternative-approach}. All of these methods lead to consistent resonance conditions.


\textbf{Harmonic Magnetic Background~~}
We now turn to a scenario where the orientation is fixed but its amplitude varies along the propagation direction { as examined by Ref.~\cite{Sikivie:1983ip,VanBibber:1987rq,Arias:2010bha,Arias:2010bh}.}
{Since the $\gamma_\perp$ state perpendicular to the background magnetic field now remains inert and unaffected by oscillations, we take the following two-component EOM}:
\begin{align}
  \label{eq:harmonic-magnetic-field}
&   i \frac{\partial }{\partial z}
  \begin{bmatrix}
    \gamma_\parallel \\ a
  \end{bmatrix}
   =
    \frac{1}{2\omega} H_{\rm A}^2(\theta)
    \begin{bmatrix}
      \gamma_\parallel\\
      a
    \end{bmatrix}
    \\
  & =
    \frac{1}{2\omega}
    \begin{bmatrix}
    0 & g_{a\gamma} \omega B\,{\rm c}_\theta\\
    g_{a\gamma}  \omega B \,{\rm c}_\theta  & \Delta m^2_{a\gamma}
    \end{bmatrix}
    \begin{bmatrix}
      \gamma_\parallel\\
      a
    \end{bmatrix}\, .
    \notag
\end{align}
The diagonal $\Delta m_{a\gamma}^2/2\omega$ term can be factored out by $a \rightarrow e^{-i(\Delta m_{a\gamma}^2/2\omega)z}a$,
then the equation above is rewritten as
\begin{align}
\label{eq:ModEoMHarm}
&   i \frac{\partial }{\partial z}
  \begin{bmatrix}
     \gamma_\parallel \\ a
  \end{bmatrix} \\
   = &
    \begin{bmatrix}
      0 & \frac{g_{a\gamma} B}{2}\,{\rm c}_\theta \,e^{-i (\Delta m_{a\gamma}^2/2\omega)z}\\
      \frac{g_{a\gamma} B}{2}\,{\rm c}_\theta \,e^{i (\Delta m_{a\gamma}^2/2\omega)z}  & 0
    \end{bmatrix}
    \begin{bmatrix}
      \gamma_\parallel\\
      a
    \end{bmatrix}\, .
    \notag
\end{align}
In the case where the frequency of $\theta$ variation exactly matches with $\Delta m_{a\gamma}^2/2\omega$, denoted as $|\dot{\theta}|=\Delta m_{a\gamma}^2/2\omega$, the off-diagonal term  retains a constant value of $g_{a\gamma}B/4$.
This constant term accumulates the oscillation phase with the additional factor of $1/2$ stemming from $\cos\theta = (e^{i\theta}+e^{-i\theta})/2$; {the other fast-oscillating term with $2\theta$ would be cyclic-averaged to effectively vanish.}
This specific condition corresponds to a parametric resonance with respect to the strength of the background magnetic field.
As in a rotating magnetic field profile, the resonance phenomenon arises when the frequency of $\theta$ variation matches the characteristic frequency associated with the a-$\gamma$ system.
Consequently, the oscillations experience significant amplification, leading to an enhanced conversion probability. The conversion probability based on Eq.~\eqref{eq:ModEoMHarm} is compared with the numerical result in Fig.~\ref{fig:num-vs-ana}.

We show in Appendix~\ref{sec:harmonic-vs-helical} 
an alternative approach where the harmonic magnetic profile is decomposed into two helical modes. We find the same result.

\begin{figure}
    \centering
    \includegraphics[width=.49\textwidth]{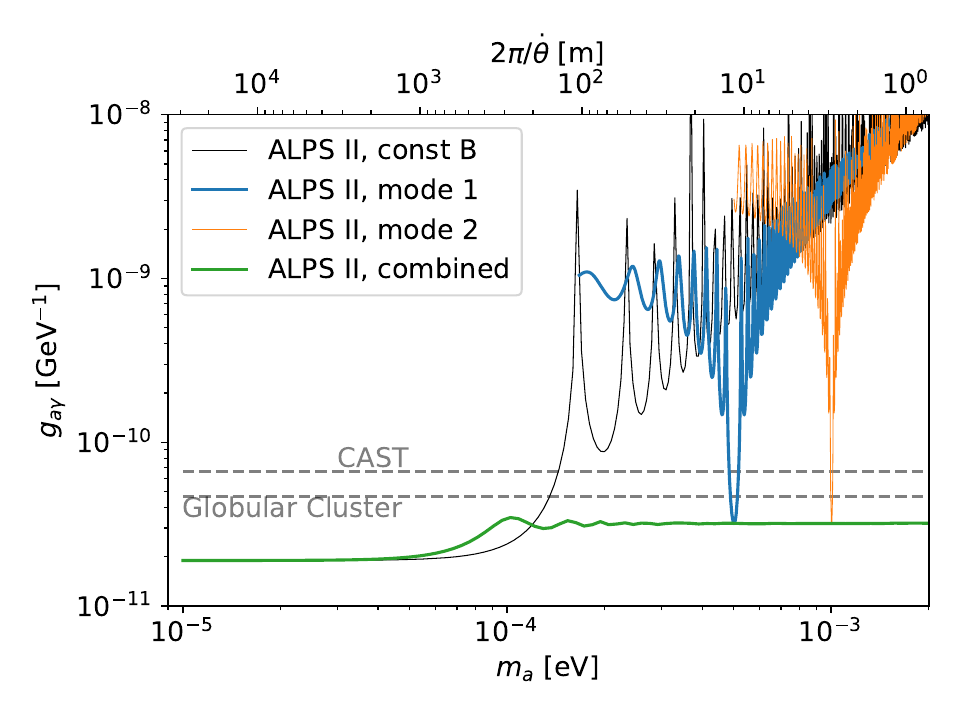}
    \caption{We show the ALPS II projection based on \cite{Ortiz:2020tgs,Bahre:2013ywa} in black. Two benchmarks {for a rotating magnetic field are achieved by fixing $\dot\theta=m_a^2/2\omega$} at $m_a=5\times 10^{-4}\,{\rm eV}$ (blue) and $m_a=10^{-3}\,{\rm eV}$ (orange). By scanning over $\dot\theta$ we achieve the improved contour (green). Note that we assume a linearly polarized photon beam. In the combined contour, the linear regime $m_a \lesssim 10^{-4}\;\mathrm{eV}$ depends on the photon linear polarization direction, while the higher mass region is $2^{3/4}$ higher than the original ALPS II plateau regardless of the photon linear polarization direction. 
    For comparison, CAST~\cite{CAST:2017uph} and globular cluster~\cite{Dolan:2022kul} are shown in grey dashed lines.}
    \label{fig:ma-ga-perfect}
\end{figure}

\textbf{Experimental Applications~~}
We take ALPS II as an example and show the amount of improvement we expect at $m_a \gtrsim 10^{-4}~\mathrm{eV}$.
We show in Fig.~\ref{fig:ma-ga-perfect} the reach in $g_{a\gamma}$ by fixing 
$\dot\theta$ to two different realistic values. 
{
}
%

Interestingly, the Relativistic Heavy Ion Collider (RHIC) at Brookhaven National Laboratory (BNL) uses helical dipole magnets to collide polarized protons beams~\cite{zotero-21781,zotero-21778,zotero-21777,Anerella:2003se}. 
The technology of magnet assembly in 2003 reached $4\,\mathrm{T}$ dipole field that rotates $2\pi$ in a length of $2.4~\mathrm{m}$~\cite{Anerella:2003se}. 
{
{

Since ALPS II utilizes an optical cavity to enhance both the laser power and the detectivity, one may be worried about whether the enhancement we discuss applies to the optical cavity photons. We show in Appendix~\ref{sec:optical-cavity} that this is indeed the case: all the analysis can be carried over for the cavity photons. 

Aside from a helical magnet profile design, we note that it is also possible to achieve equivalent enhancement through the modulation of lasers. In Appendix~\ref{sec:additional-exp-setup} we show extra methods of modulating the laser beam that can achieve comparable enhancement. 

{
\textbf{Scanning Method}~~ 
As we have shown, the resonance condition is $m_a^2/2\omega = |\dot \theta|$. We conclude that a given helical setup can only cover axion masses in a narrow range determined by the resonance width. One, therefore, needs to adjust the helical frequency, $\dot\theta$, to scan over different axion masses. In Ref.~\cite{Mun:2014zva} it is shown that $\sim 10\%$ adjustment can be achieved in helical undulators.

To utilize the resonance across a range of axion masses, we now relax the requirement on the helical magnetic field profile with a new approach. Instead of helical undulators, we will adopt dipole modules. The ``helicity'' is created by arranging each magnet to have an axial angle $\Delta \theta$ relative to the one preceding it, as illustrated in Fig.~\ref{fig:discrete-helical}. By adjusting the relative orientation between magnets in each run, we can cover a significant range of the axion parameter space. 

Importantly, apart from the mechanical effort, ALPS II can continue utilizing the HERA magnets currently in use, instead of repurposing the RHIC helical magnets. This will significantly reduce experimental costs and make adjustments much easier.

\begin{figure}
    \centering
    \includegraphics[width=.5\textwidth]{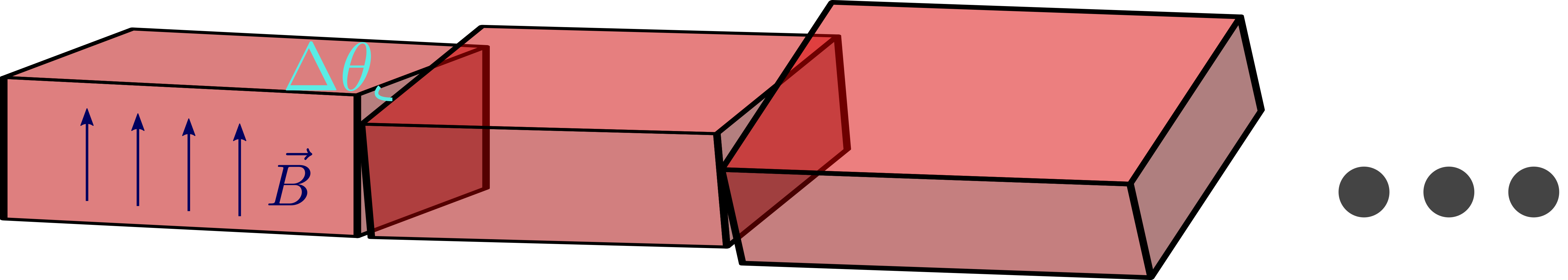}
    \caption{An illustration of the discrete helical array setup. }
    \label{fig:discrete-helical}
\end{figure}
We will use the HERA magnet specifications (B=5.3~T, with a length of $ 8.83\,\mathrm{m}$) in what follows to compute its sensitivity reach. We note that the discrete helical array is a generalization of the original Wiggler proposal~\cite{VanBibber:1987rq}, which is a special case of setting $\Delta \theta = \pi$. 
The results are shown in Fig.~\ref{fig:discrete-helical-reach}. A particularly interesting feature of the discrete helical setup is that resonance can happen at certain $m_a^2 \gg 2\omega \dot\theta$; these higher integer resonance modes are discussed in Appendix~\ref{sec:high-modes}.
We will comment more on the comparison with the Wiggler method in Appendix~\ref{sec:comparison}.

\begin{figure}
    \centering
    \includegraphics[width=.5\textwidth]{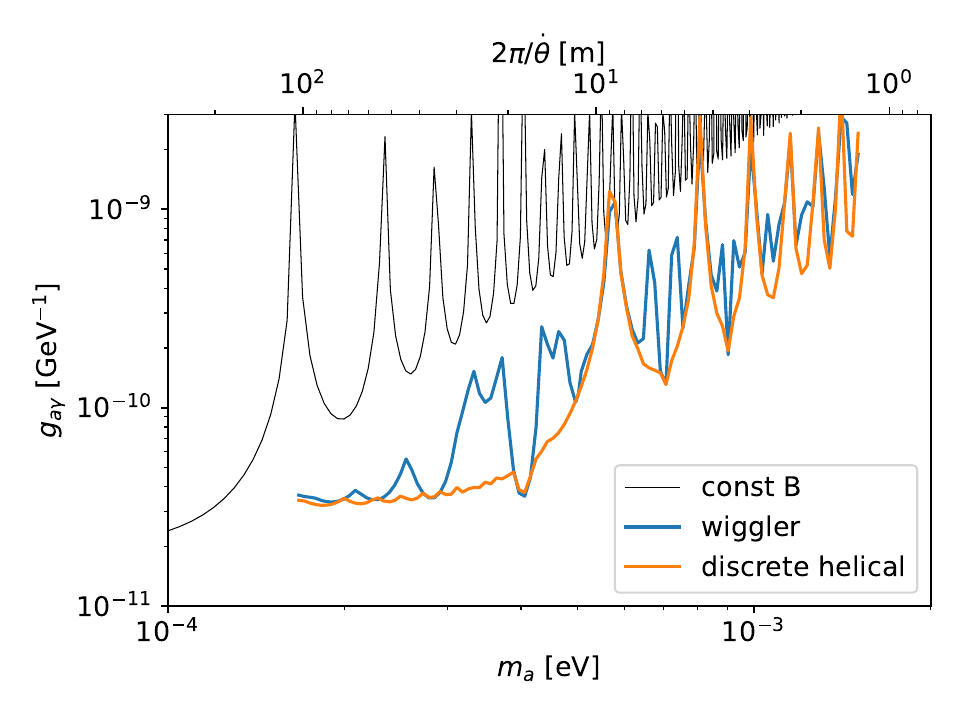}
    \caption{The experimental reach of discrete helical array and the original Wiggler method using the existing magnets at ALPS II (HERA magnets). We follow \cite{VanBibber:1987rq} and perform 15 configurations for $\Delta \theta=\pi/n$ with $n\in \{1, 2,...,15\}$ (orange). The Wiggler setup is the special case of $n=1$ (blue).}
    \label{fig:discrete-helical-reach}
\end{figure}

}


\textbf{Conclusion~~} In the {small-mixing non-linear regime} of axion-photon conversion, the conversion probability is independent of the baseline length but suppressed by the axion mass. Put differently, once the axion-photon oscillation enters the non-linear regime, not the full length of the conversion baseline is fully used.
We propose an experimental setup to alleviate this limitation with a spatially varying magnetic field. We showed that one way to think about the effect of a varying magnetic field is that, with a carefully chosen profile, its spatial variation can lower the axion oscillation wave number, which increases the conversion amplitude (\textit{i.e.} the prefactor in Eq.~(\ref{eq:conversion-prob-varyingB})). As a result, it delays the onset of the nonlinear regime and makes better use of the full baseline.

{
We note that this type of AMR is a great way to pinpoint the axion mass should any anomalies arise (\textit{e.g.} from BabyIAXO or astrophysical sources)\footnote{This was first suggested by Andreas Ringwald through private communications.}. 
}

{

As the signal photons will have a rotating polarization vector in the rotating $\mathbf B$ setup, it can potentially open up new experimental searching strategies. Although the scanning of axion mass will require longer operation time, we stress that this has great advantage over simply increasing the running time with a constant $\mathbf{B}$ field, where the reach in $g_{a\gamma}$ can only increase as (time)$^{1/8}$ at best.

We highlight the relation and difference with previous work~\cite{Sikivie:1983ip,VanBibber:1987rq,Jaeckel:2007gk,Arias:2010bha,Arias:2010bh,Wang:2015dil,Zarei:2019sva,Sharifian:2021vsg}, where the magnetic field or the photon (electric field) is manipulated to enhance the experimental sensitivity in the heavy axion regime, in more detail in Appendix~\ref{sec:comparison}. 
The code that reproduces all results is released at \href{https://github.com/ChenSun-Phys/axion-magnetic-resonance}{GitHub}.

~\\
\textit{Acknowledgement}~~~~ We would like to thank Daniele Alves for useful comments on unitarity and the enhancement related to the during of the experiment; Michael Graesser for the effect of an induced electric field; Andrea Albert, James Colgan, Leanne Delma Duffy, Marianne Francois, Kranti Gunthoti, Christopher Lee, Xuan Li, Ming Xiong Liu, Charles Alan Rohde, Adam Lewis Susser, and Janardan Upadhyay for comments on the technical aspect of the experimental setup; Weiyao Ke for discussions on the helical magnet; Tomer Volansky for the access to Tel Aviv Univeristy High-Performance Computing, where part of the work was performed. We also thank Axel Lindner, Gilad Perez, Andreas Ringwald, and Nicholas~Rodd for useful communications.  
{
HS is supported by the Deutsche Forschungsgemeinschaft under
Germany Excellence Strategy — EXC 2121 “Quantum
Universe” — 390833306. 
 This work was supported by the U.S. Department of Energy through the Los Alamos National Laboratory. Los Alamos National Laboratory is operated by Triad National Security, LLC, for the National Nuclear Security Administration of U.S. Department of Energy (Contract No. 89233218CNA000001). Research presented in this article was supported by the Laboratory Directed Research and Development (LDRD) program of Los Alamos National Laboratory under projects 20220135DR and 20230047DR.
SY was supported by IBS under the project code, IBS-R018-D1.
CS and SY acknowledge the Galileo Galilei Institute (GGI) for their generous hospitality during the workshop “New Physics from the Sky,” where this work was initiated.

\vfill
\pagebreak
~~
\vfill
\pagebreak

\appendix
\begin{widetext}
    \begin{center}
        {\Large \bf Supplemental Material}
    \end{center}
~\\~\\      
\end{widetext}

\section{Axion-Photon Dynamics in a Constant Magnetic Field}
\label{sec:axion-photon-usual-formula}

The equation of motion in the relativistic limit is given by $i \partial _z \tilde \Psi = \frac{H^2}{2\omega} \tilde \Psi$. In components, it reads~\cite{Raffelt:1987im,Mirizzi:2006zy,Mirizzi:2009nq}
\begin{align}
  \label{eq:old-eom}
  i \frac{\partial }{\partial z}
  \begin{bmatrix}
    \gamma_\perp \\ \gamma_\parallel \\ a
  \end{bmatrix}
  & =
    \frac{1}{2\omega}
    \begin{bmatrix}
      m_\gamma^2 & 0 & 0 \\
      0 & m_\gamma^2 & g_{a\gamma}  \omega B \\
      0 & g_{a\gamma} \omega B  & m_a^2
    \end{bmatrix}
    \begin{bmatrix}
      \gamma_\perp \\
      \gamma_\parallel\\
      a
    \end{bmatrix}\, .
\end{align}

In the conventional setup of LSTW experiments, $\dot{\theta} = 0$, the $a$-to-$\gamma$ conversion probability is given by
\begin{align}
  \label{eq:conversion-probability}
{P_{a{\rightarrow} \gamma_\parallel}}
& = 
\frac{(g_{a\gamma}B)^2}{\Delta_{\rm osc}^2}
  \sin^2 \left (\frac{\Delta_{\rm osc} l}{2} \right)
  \\
  & 
  \approx
  \begin{cases}
     (g_{a\gamma} B l)^2/4 & ~~\Delta_{\rm osc} l \ll 1 \,,\\
    (g_{a\gamma}B)^2/2 \Delta_{\rm osc}^2 & ~~\Delta_{\rm osc} l \gg 1 \,,
  \end{cases}
    \notag
\end{align}
where $\Delta_{\rm osc} = \sqrt{(\Delta m_{a\gamma}^2/2\omega)^2+(g_{a\gamma} B)^2}$, and $l$ denotes the domain length of the background magnetic field.
Ignoring the photon refractive index $m_\gamma = 0$, the conversion probability becomes independent of the axion mass if the domain length is sufficiently short to impose a small phase retardation condition $\Delta_{\rm osc} l \ll 1$; this is dubbed as the ``linear'' regime.
On the other hand, when the oscillation length $2\pi\Delta_{\rm osc}^{-1}$ is significantly shorter than the domain length as the so-called ``non-linear'' regime, {a rapid oscillation of conversion probability is averaged out over fluctuations of the parameters, e.g., $\omega$ and $l$\,,}
and the conversion probability is controlled by the mixing angle, which is suppressed by the axion mass squared {in the small-mixing limit, $m_a^2/2\omega \gg g_{a\gamma} B$}.

\section{Initial Photon Polarization}
\label{sec:appendix-ic}
The production of axions from photon initial states has a dependence on the photon polarization. We further make some clarifications here and discuss the applicability of Eq.~\eqref{eq:conversion-prob-varyingB}.

Let us start with Eq.~\eqref{eq:explicit-show-resonance-helical} and look at the case $\dot \theta = m_a^2/2\omega$. We always assume {a small axion-photon mixing}, $g_{a\gamma} B \ll m_a^2/\omega$. In the linear regime, we have $z \, m_a^2/\omega \lesssim 1$. Therefore, the conversion amplitude between $\Psi_{1}$ and $a$ is approximated by
\begin{widetext}
\begin{align}
\mathcal M_{1a}
& \approx 
\begin{bmatrix}
    0 & 0 & 1
\end{bmatrix}
\left (
\mathbb{1} - 
i \begin{bmatrix}
-\dot\theta & 0 & g_{a\gamma}B/2\sqrt{2}\\
0 & \dot\theta & g_{a\gamma}B/2\sqrt{2} \\
g_{a\gamma}B/2\sqrt{2} & g_{a\gamma}B/2\sqrt{2} & \dot\theta
\end{bmatrix} {z}
\right )
\begin{bmatrix}
    1 \\ 0 \\ 0
\end{bmatrix}
\approx 
\frac{-i}{2\sqrt 2} g_{a\gamma} B \, z.
\\
\mathcal M_{2a}
& \approx 
\begin{bmatrix}
    0 & 0 & 1
\end{bmatrix}
\left (
\mathbb{1} - 
i \begin{bmatrix}
-\dot\theta & 0 & g_{a\gamma}B/2\sqrt{2}\\
0 & \dot\theta & g_{a\gamma}B/2\sqrt{2} \\
g_{a\gamma}B/2\sqrt{2} & g_{a\gamma}B/2\sqrt{2} & \dot\theta
\end{bmatrix}{z}
\right )
\begin{bmatrix}
    0 \\ 1 \\ 0
\end{bmatrix}
\approx 
\frac{-i}{2\sqrt 2} g_{a\gamma} B \, z.
\end{align}
\end{widetext}
If we start with a linear initial state $[1 ~ 0 ~ 0]^T$ in the $[\gamma_\perp ~ \gamma_\parallel ~ a]^T$ lab frame, after the $V$ rotation as shown in Eq.~\eqref{eq:diagonalize-12}, we have $\Psi = [-i/\sqrt{2}  ~ i/\sqrt{2} ~ 0]^T$. The amplitudes between $\Psi_1$ component converting to axion and $\Psi_2$ to axion will interfere destructively, as $\mathcal M_{1a}/\sqrt{2} - \mathcal M_{2a}/\sqrt{2}\approx 0$. This is indeed what we observe numerically. 

If, on the other hand, we start with $[ 0 ~ 1 ~ 0]^T$ initial state lab frame, we will overcome this destructive interference. In the rotating basis we have $\Psi = [1/\sqrt 2  ~ 1/\sqrt 2 ~ 0]^T$. The axion appearance probability is given by $P_{\gamma a} = |\mathcal M_{1a}/\sqrt{2} + \mathcal M_{2a}/\sqrt{2}|^2 \approx \frac{1}{4} (g_{a\gamma}B z)^2$, which is the same as Eq.~\eqref{eq:conversion-prob-varyingB} in the linear regime. 

In the nonlinear regime, the two amplitudes have very different magnitudes and phases. In our example, $\mathcal M_{1a} \sim g_{a\gamma}B \omega/\Delta m_{a\gamma}^2$, while $\mathcal M_{2a} \sim \mathcal O(1)$ {$[i \sin\left(g_{a\gamma}Bz/2\sqrt{2}\right)]$} .  Therefore, the aforementioned two choices in the initial states will lead to the same axion production rate $P_{\gamma a} \approx |\mathcal M_{2a}/\sqrt{2}|^2 \approx P_{a\gamma}/2$. 

This is of relevance to the experimental setup as we show that the dependence of the axion production only relies on the initial photon polarization in the linear regime. Since our main improvement is the non-linear regime -- for ALPS II this corresponds to $m_a \gtrsim 10^{-4}$ -- the result is not affected by the initial photon polarization relative to the magnetic field direction.

\section{Alternative Approach -- Parametric Resonance}
\label{sec:parametric}
{
We show that the AMR with a helical magnetic field can also be understood using the parametric resonance language. Let us again start from the EOM:
\begin{align}
  &   i \partial_z
    \tilde \Psi
   =
    \frac{1}{2\omega} H^2(\theta)
    \tilde \Psi    
    \\
  & =
    \frac{1}{2\omega}
    \begin{bmatrix}
      0 & 0 & g_{a\gamma}  \omega B \,{\rm s}_\theta \\
      0 & 0 & g_{a\gamma} \omega B \,{\rm c}_\theta\\
      g_{a\gamma} \omega B \,{\rm s}_\theta & g_{a\gamma}  \omega B \,{\rm c}_\theta  & \Delta m^2_{a\gamma}
    \end{bmatrix}
    \begin{bmatrix}
      \gamma_\perp \\
      \gamma_\parallel\\
      a
    \end{bmatrix},
    \notag
\end{align}
After factorizing out the fast oscillation of $a$, we get
\begin{widetext}
\begin{align}
\label{eq:parametric-resonance-helical}
     i \partial_z
    \tilde \Psi
  & \approx
    \frac{1}{2\omega}
    \begin{bmatrix}
      0 & 0 & g_{a\gamma}  \omega B \,{\rm s}_\theta\,e^{{-}i (\Delta m_{a\gamma}^2/2\omega)z} \\
      0 & 0 & g_{a\gamma} \omega B \,{\rm c}_\theta\,e^{{-}i (\Delta m_{a\gamma}^2/2\omega)z}\\
      g_{a\gamma} \omega B \,{\rm s}_\theta\,e^{{+}i (\Delta m_{a\gamma}^2/2\omega)z} & g_{a\gamma}  \omega B \,{\rm c}_\theta  \,e^{{+}i (\Delta m_{a\gamma}^2/2\omega)z} & 0
    \end{bmatrix}
    \begin{bmatrix}
      \gamma_\perp \\
      \gamma_\parallel\\
      a
    \end{bmatrix}.
    \notag
\end{align}
\end{widetext}
The off-diagonal term has an oscillation that is faster than the $\gamma-a$ mixing caused by $g_{a\gamma}\omega B$, where it can be cycle-averaged to zero, unless $\dot\theta = \pm \Delta m_{a\gamma}^2/2\omega$, in which case we have a constant mixing term. 
}

\section{Alternative Approach -- Dispersion Relation and Reference Frame}
\label{sec:alternative-approach}
\begin{figure*}[ht]
    \centering
    \includegraphics[width=.7\textwidth]{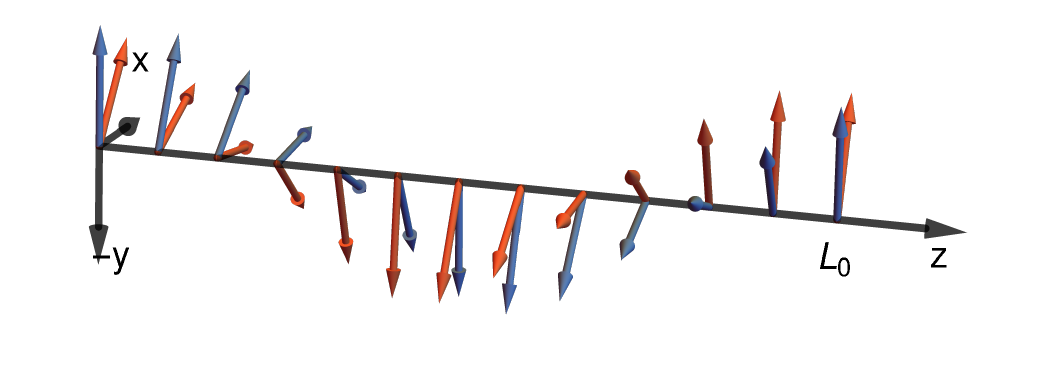}
    \includegraphics[width=.7\textwidth]{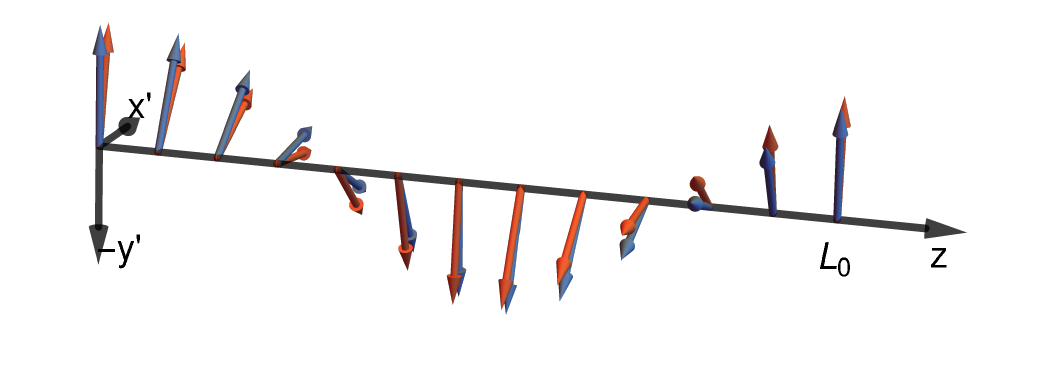}    
    \caption{Cartoon demonstrating the observed rotation of the photon polarization vector in the inertial lab frame (upper panel) and a non-inertial frame that rotates with frequency $\dot\theta$ with respect to the lab (lower panel). The colored arrows represent the polarization vector. The blue ones are from the snapshot taken at $t=0$, red ones are the snapshot at $t=\Delta t$. In the lower panel, because the reference frame is rotating in the same direction as the polarization vector, the observed phase after the same amount of time $\Delta t$ is smaller, as represented by the angle between the red and blue arrows. However, since there the phase dependence in the $z$ direction is not affected, one can find the same phase at $z=0$ and $z=L_0 (= \lambda)$ regardless of the reference frame.}
    \label{fig:inertial-frame-vs-non-inertial-frame}
\end{figure*}
{The key point of the axion magnetic resonance is to modify the photon dispersion relation effectively such that it matches the axion dispersion relation.}
Let us verify this using a more heuristic approach. Since $\dot\theta \ll \omega$, we take the classical picture to illustrate the effect. 

We start with the inertial frame of the lab. Inside a vacuum, the photon dispersion relation tells us $\omega = k$. Without loss of generality, let us look at circularly polarized photons that have helicity $-1$. The polarization vector of the gauge field at each point rotates with the frequency $\omega$. 

Suppose the magnetic field rotates in the same direction as the photon helicity with a frequency of $\dot\theta$. Now, let us go to the frame where the magnetic field constantly points to the $y$ direction. In this frame, an observer will find the photons with a frequency $\omega' = \omega - \dot\theta$, while $k$ stays the same. 

This can be understood as follows. At each point, because the observer is rotating in the same direction as does the photon polarization vector, to the observer the rotation becomes slower, \textit{i.e.} it takes longer time to complete the rotation of $2\pi$. As a result, $\omega' < \omega$. On the other hand, since there is no transformation in the $z$ direction, one can still find the same distance between nearest points where the same phase is shared, regardless of being in the rotating frame or the lab frame, \textit{i.e.} $k'=k$. 
A cartoon demonstrating this point is shown in Fig.~\ref{fig:inertial-frame-vs-non-inertial-frame}.

One might get the impression that only a mere change of reference frame would do the job while no rotating magnetic field is needed. That is not the case. 
Let us go back to the static magnetic field along the $y$ axis of the lab frame. If we repeat the change of frame \textit{without} an actual rotating magnetic field,  Eqs.~\eqref{eq:U-rot}-\eqref{eq:diagonalize-12} gives us
\begin{align}
  i \partial _z
  \Psi
  & = 
    \begin{bmatrix}
      - \dot \theta & 0 & \frac{g_{a\gamma}  B}{2\sqrt 2}\mathrm{e}^{+i\theta} \\
      0 & \dot \theta &  \frac{g_{a\gamma}  B}{2\sqrt 2} \mathrm{e}^{-i\theta} \\
        \frac{g_{a\gamma}  B}{2\sqrt 2}\mathrm{e}^{-i\theta} & \frac{g_{a\gamma}  B}{2\sqrt 2} \mathrm{e}^{+i\theta}  & \frac{\Delta m_{a\gamma}^2}{2\omega} 
    \end{bmatrix}
    \Psi \, .
\end{align}
{After factorizing out the phase $\mathrm{e}^{i\dot\theta \, z}$, $\mathrm{e}^{-i\dot\theta \, z}$, and $\mathrm{e}^{-i\Delta m^2_{a\gamma}/2\omega}$, the mixing term only has an oscillation of frequency $\Delta m^2_{a\gamma}/2\omega$: 
\begin{align}
  i \partial _z
  \Psi
  & = 
    \begin{bmatrix}
      0 & 0 & \frac{g_{a\gamma}  B}{2\sqrt 2}\mathrm{e}^{-i \frac{\Delta m_{a\gamma}^2}{2\omega} z} \\
      0 & 0 &  \frac{g_{a\gamma}  B}{2\sqrt 2} \mathrm{e}^{-i\frac{\Delta m_{a\gamma}^2}{2\omega} z} \\
        \frac{g_{a\gamma}  B}{2\sqrt 2}\mathrm{e}^{+i\frac{\Delta m_{a\gamma}^2}{2\omega} z} & \frac{g_{a\gamma}  B}{2\sqrt 2} \mathrm{e}^{+i\frac{\Delta m_{a\gamma}^2}{2\omega} z}  & 0
    \end{bmatrix}
    \Psi \, ,
\end{align}
which is independent of $\dot\theta$, the parameter we used for the change of reference frame that otherwise does not correspond to any physical quantities. In the small-mixing non-linear regime, \textit{i.e.} $g_{a\gamma} B\ll \Delta m_{a\gamma}^2/2\omega$, the mixing term is averaged out, which leads to  the usual suppression of the conversion probability. 
}

Put differently, the rotating magnetic field chooses a preferred frame, in which the photon dispersion relation is modified to be the same as that of the axion. {This is somewhat similar to the method we used in Eq.~\eqref{eq:U-rot}, where choosing a given helical profile $\theta(z)$ is equivalent to fixing the gauge. }

\section{Harmonic vs Helical}
\label{sec:harmonic-vs-helical}
We can understand the resonant enhancement by a harmonic magnetic profile in terms of two helical magnetic fields.
We decompose the harmonic magnetic field into two helical magnetic fields with opposite rotating directions. Let us denote the Hamiltonian with a helical profile in Eq.~(\ref{eq:rotating-B-int-basis}) to be {$H_E^2(\theta; ~g_{a\gamma} B, \Delta m_{a\gamma}^2)$. We can decompose $H_A^2$ into}
\begin{align}
  \label{eq:decomposition}
  H_A^2 =   H_E^2(\theta; \frac{g_{a\gamma} B}{2}, \Delta m_{a\gamma}^2)  + H_E^2(-\theta; \frac{g_{a\gamma} B}{2}, 0),
\end{align}
where we used the shorthand $\theta$ but both $H_A^2$ and $H_E^2$ depend on $\theta(z)$ explicitly. 
Now we can treat the first term the same way as we did with Eq.~\eqref{eq:rotating-B-int-basis}. {
The resulted EOM reads
\begin{align}
  \label{eq:harmonic-extra-term}
  i \partial _z
  \Psi
  & = 
    \begin{bmatrix}
      - \dot \theta & 0 & \frac{g_{a\gamma}  B/2}{2\sqrt 2} \\
      0 & \dot \theta &  \frac{g_{a\gamma}  B/2}{2\sqrt 2} \\
        \frac{g_{a\gamma}  B/2}{2\sqrt 2} & \frac{g_{a\gamma}  B/2}{2\sqrt 2}  & \frac{\Delta m_{a\gamma}^2}{2\omega} 
    \end{bmatrix} \Psi
\\ 
& +     \begin{bmatrix}
      0 & 0 & \frac{g_{a\gamma}  B/2}{2\sqrt 2} \,\mathrm{e}^{+i2\theta} \\
      0 & 0 &  \frac{g_{a\gamma}  B/2}{2\sqrt 2} 
      \,\mathrm{e}^{-i2\theta} \\
        \frac{g_{a\gamma}  B/2}{2\sqrt 2} \,\mathrm{e}^{-i2\theta}& \frac{g_{a\gamma}  B/2}{2\sqrt 2} 
        \,\mathrm{e}^{+i2\theta}& 0
    \end{bmatrix}
    \Psi \, .  
    \notag
\end{align}
}
{
We can factorize out the phase $\mathrm{e}^{+i\theta}$, $\mathrm{e}^{-i\theta}$, $\mathrm{e}^{-i \Delta m_{a\gamma}^2/2\omega}$, after which the EOM reads
\begin{align}
  i \partial _z
  \Psi
  & = 
    \begin{bmatrix}
      0 & 0 & \frac{g_{a\gamma}  B/2}{2\sqrt 2} \mathrm{e}^{-i \Omega_+\, z} \\
      0 & 0 & \frac{g_{a\gamma}  B/2}{2\sqrt 2} \mathrm{e}^{-i \Omega_-\, z}  \\
       \frac{g_{a\gamma}  B/2}{2\sqrt 2} \mathrm{e}^{+i \Omega_+\, z}  & \frac{g_{a\gamma}  B/2}{2\sqrt 2} \mathrm{e}^{+i \Omega_-\, z} & 0
    \end{bmatrix} \Psi
\cr
& +     \begin{bmatrix}
      0 & 0 & \frac{g_{a\gamma}  B/2}{2\sqrt 2} \,\mathrm{e}^{-i \Omega_-\, z} \\
      0 & 0 &  \frac{g_{a\gamma}  B/2}{2\sqrt 2} 
      \,\mathrm{e}^{-i \Omega_+\, z} \\
        \frac{g_{a\gamma}  B/2}{2\sqrt 2} \,\mathrm{e}^{+i \Omega_-\, z}& \frac{g_{a\gamma}  B/2}{2\sqrt 2} 
        \,\mathrm{e}^{+i \Omega_+\, z}& 0
    \end{bmatrix}
    \Psi \, ,
\end{align}
where $\Omega_{\pm} = \Delta m^2_{a\gamma}/2\omega  \pm \dot \theta $. Without loss of generality, we choose $\Omega_-=\Delta m_{a\gamma}^2/2\omega - \dot\theta = 0, \Omega_+ = \Delta m_{a\gamma}^2/\omega$. Therefore, the EOM can be rearranged as 
\begin{align}
\label{eq:harmonic-decompose-helical}
  i \partial _z
  \Psi
  & = 
    \begin{bmatrix}
      0 & 0 & \frac{g_{a\gamma}  B/2}{2\sqrt 2}  \\
      0 & 0 & \frac{g_{a\gamma}  B/2}{2\sqrt 2}  \\
       \frac{g_{a\gamma}  B/2}{2\sqrt 2}  & \frac{g_{a\gamma}  B/2}{2\sqrt 2}  & 0
    \end{bmatrix} \Psi
\\
& +     \begin{bmatrix}
      0 & 0 & \frac{g_{a\gamma}  B/2}{2\sqrt 2} \,\mathrm{e}^{-i \Omega_+\, z} \\
      0 & 0 &  \frac{g_{a\gamma}  B/2}{2\sqrt 2} 
      \,\mathrm{e}^{-i \Omega_+\, z} \\
        \frac{g_{a\gamma}  B/2}{2\sqrt 2} \,\mathrm{e}^{+i \Omega_+\, z}& \frac{g_{a\gamma}  B/2}{2\sqrt 2} 
        \,\mathrm{e}^{+i \Omega_+\, z}& 0
    \end{bmatrix}
    \Psi \, ,
    \notag
\end{align}
In the small-mixing nonlinear regime, we have $\Delta m_{a\gamma}^2/2\omega \gg g_{a\gamma}B$. Therefore, the second term can be safely averaged out. This is the same as Eq.~\eqref{eq:ModEoMHarm} up to a $45^\circ$ rotation. 
}

\section{Noise in the Magnetic Helical Profile}
\label{sec:noise-discussion}
All the above assumes a constant rotating frequency of the magnetic field. Next, let us demonstrate the robustness of AMR against noises in the magnetic frequency. 
We take the fluctuation of the rotation frequency to be within a certain value of the central value, \textit{e.g.} $1\;\%$, or $10\;\%$. More precisely, we assume
\begin{align}
    \dot\theta = \bar {\dot\theta} + \delta \dot\theta\,,
\end{align}
where $\delta \equiv \delta \dot\theta/ \bar {\dot\theta} $ is a Gaussian random variable centered around zero, with a standard deviation of 0.01 or 0.1. 

{
Let us further define the frequency of the noise as
\begin{align}    
\omega_\delta 
& \equiv 
\frac{1}{\delta\dot\theta} \frac{d \delta\dot\theta}{dz}.
\end{align}
When the noise frequency is very low, \textit{i.e.} $\omega_\delta\ll (100~\mathrm{m})^{-1}$, $\dot \theta$ can be treated as constant throughout the propagation of most of the whole baseline: the matched $m_a$ is simply not the one we expect, $m_a^2 = 2\omega \bar{\dot\theta}$, but rather $m_a^2 = 2\omega (\bar{\dot\theta}+\delta\dot\theta)$. The reach in $g_{a\gamma}$ is otherwise not impacted. 

On the other hand, when the noise is at high frequency, the effect is likely to average out as we demonstrate in Fig.~\ref{fig:num-vs-ana-1pct}. Therefore, the most damaging noise comes from $1\,\mathrm{m} \lesssim \omega_\delta^{-1} \lesssim 100\,\mathrm{m}$.

We take two benchmarks of the noise frequency, 28~MHz (corresponding to $\delta$ change 10 times during the propagation of the 106~meter baseline) and 5.7~GHz (corresponding to 2000 changes of $\delta$). We approximate that $\delta$ stays constant after each time it changes to a different value until the next change. The values of $\delta$ in different time intervals are uncorrelated. 
We make $100$ polls of the Gaussian distribution of $\delta$ for each $m_a$ point. We show in Fig.~\ref{fig:gaussian-noise} the averaged contour of the repeated $100$ scans in $m_a$ as well as the $2\,\sigma$ uncertainty band of the contour. We observe that great improvement in ALPS II can still be achieved even with this very conservative assumption on the regularity of the helical profile.
\begin{figure}[th]
  \centering
  \includegraphics[width=.49\textwidth]{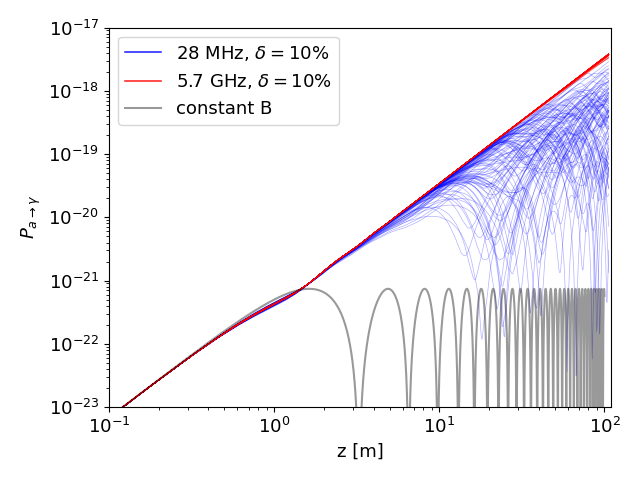}
  \caption{We take the benchmark point $m_a = 0.001\,\mathrm{eV}$, $\omega=1.16\,\mathrm{eV}~(1064\,\mathrm{nm}$), $g_{a\gamma} = 10^{-11}\,\mathrm{GeV}^{-1}$, $B = 5.3\,\mathrm{T}$, $z=106\,\mathrm{m}$. We include a 10 \% fluctuation in $\dot \theta$. We show the noise benchmarks with a fixed noise frequency 23 MHz (blue) and 5.7 GHz (red). We show 100 runs of each point. Every red and black curve corresponds to a realization of the Gaussian distributed $\dot\theta$ profile. We include the conversion probability with a constant $B$ field (grey). 
  }
  \label{fig:num-vs-ana-1pct}
\end{figure}

\begin{figure}[th]
  \centering
  \includegraphics[width=.49\textwidth]{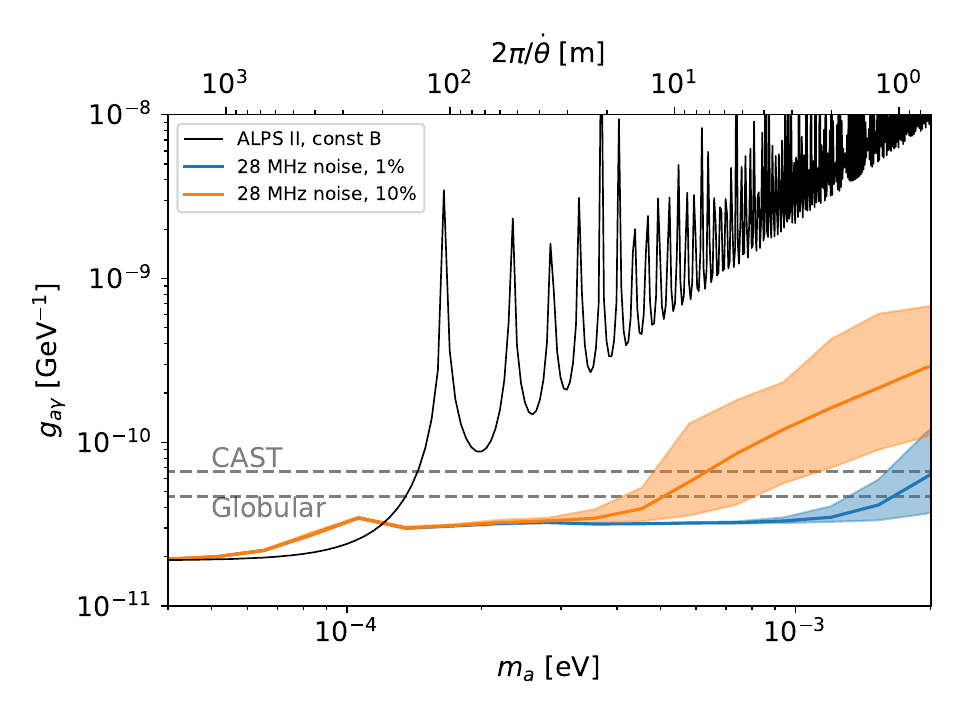}
  \includegraphics[width=.49\textwidth]{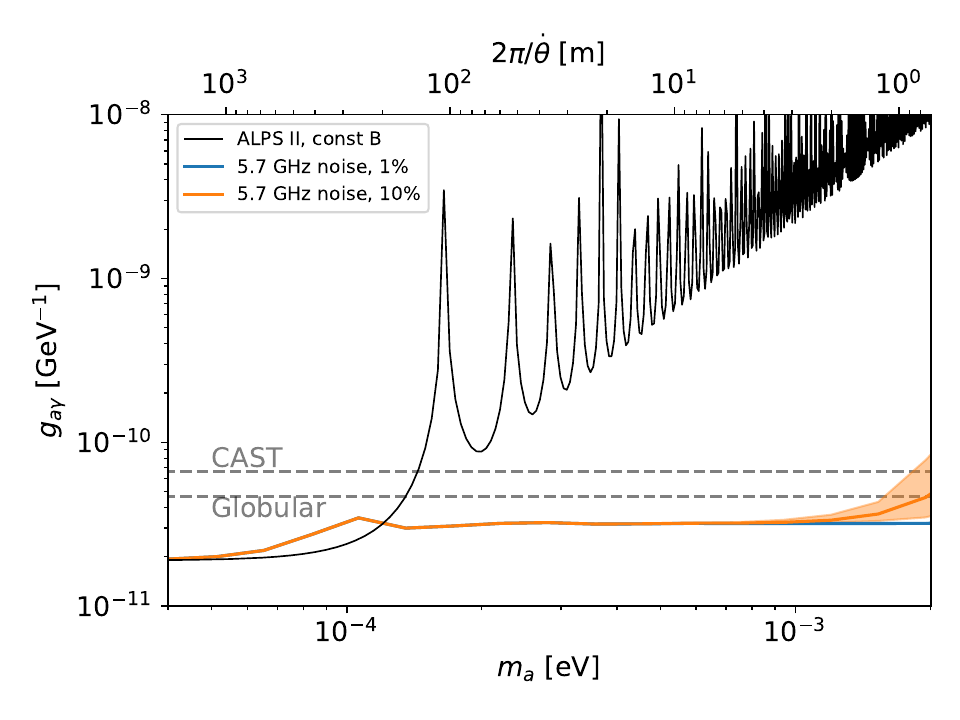}
  \caption{We add {noises} to the magnetic field rotation frequency. We make 100 scans and show the $2\,\sigma$ uncertainty of the contour due to {low}-frequency noise (upper) and {high}-frequency noise (lower). For each case, we take the Gaussian noise amplitude $\delta$ to be 1\% (blue) and 10\% (orange).
  }
  \label{fig:gaussian-noise}
\end{figure}

We show in Fig.~\ref{fig:gaussian-noise} the reach of $g_{a\gamma}$ with $1\,\%$ and $10\,\%$ irregularities in the helical profile with a noise frequency given by $\omega_\delta L \approx 10$ and $\omega_\delta L \approx 2000$. 
}

\section{The Optical Cavity}
\label{sec:optical-cavity}

In the previous computation, we studied the laser beam directly shining through the wall. In reality, ALPS~II adopts an optical cavity~\cite{Sikivie:2007qm,Bahre:2013ywa,Ortiz:2020tgs} to enhance the effective beam power. In this section, we show that the AMR at $\dot\theta=m_a^2/2\omega$ persists in the cavity setup. 

The optical cavity is designed to impose a Dirichlet boundary condition such that the photons with the following wave number are enhanced
\begin{align}
k
& = \frac{2\pi}{\lambda} = \frac{n\pi}{L},
\end{align}
where $L$ is the length of the cavity.
It is straightforward to understand the enhanced photon-to-axion conversion in the axion production cavity. For a laser with a certain wavelength, say $\lambda=1064\,\mathrm{nm}$, we can tune the cavity length by moving the mirrors such that $L = n \, \lambda/2$ for an integer $n$. This will coherently enhance the wave from the laser power $P_0$ to $P_0/\eta$, where $R=1-\eta$ is the reflectivity of the mirrors~\cite{Hoogeveen:1990vq}. Other than this enhancement of photon power, all the discussions above still hold. 

The photon regeneration, on the other hand, is trickier. {Let us start with an incident axion beam}
\begin{align}
a(x, t) = {a_0} \sin(k_a z - \omega t)\,.
\end{align}
Based on the context, we use $a(x,t)$ to denote the propagating field at $(x, t)$ while ${a_0}$ is the amplitude of the wave. 
Note that due to the cavity boundary condition only the following photons modes can stably exist:
\begin{align}
    \vec A_n(z, t) = \vec A_n(t) \sin \left (\frac{n\pi}{L} z \right ).
\end{align}
 We use $\vec A(z,t)$ and $\vec A(t)$ to denote the field and the amplitude of the wave when there is no ambiguity. 
Let us quickly review {the field dynamics in a} static magnetic field, $\mathbf B = [0~ B_0 ~ 0]^T$. 
Electromagnetic waves are sourced by the axion field with the following equation of motion.
\begin{align}
    \nabla \times \vec B - \partial_t \vec E = -g \vec B_0 \partial_t a,
\end{align}
where $B_0$ is the external magnetic field. With the assumption of no free charge and under Coulomb gauge, this can be rewritten as~\cite{Sikivie:2007qm}
\begin{align}
\label{eq:cavity-const-B}
    (\partial_t^2 + \gamma \partial_t +  \omega_n^2) 
    A_{n,y}(t) \sin \left (\omega_n z \right )
    & = 
    g  B_0 \omega {a_0} \cos (k_a z - \omega t),
\end{align}
where $\omega_n = n\pi/L$. Since there is a power leakage from the mirror, which causes the photon amplitude to decrease. This is effectively captured by the friction term $\gamma$. 
Using the orthogonality of the basis, we get
\begin{align}
    \left [\int_0^L dz\, \sin(\omega_n z)  \sin \left (\omega_n z \right )\right ]
    (\partial_t^2 + \gamma \partial_t +  \omega_n^2) 
    A_{n,y}(t) 
    \cr
    = 
    g  B_0 \omega {a_0} 
    \left [\int_0^L dz \, \sin(\omega_n z) \, \cos (k_a z - \omega t)
    \right ], 
\end{align}
which leads to 
\begin{align}
    & (\partial_t^2 + \gamma \partial_t +  \omega_n^2) 
     A_{n, y}(t) 
    \\
    & =     
    g B_0 \omega {a_0} \;
    \frac{2}{L}     
    \left [\int_0^L dz \, \sin(\omega_n z) \, \cos (k_a z - \omega t)
    \right ]
    \cr
    & =
    g  B_0 \omega {a_0} \;
    \frac{1}{L}     
    \bigg [\int_0^L dz \, \sin((k_a + \omega_n) z - \omega t) 
    \cr
    & \qquad \qquad \qquad \qquad \qquad 
    + \sin((-k_a + \omega_n) z + \omega t) 
    \bigg ].
    \notag
\end{align}
Let us adjust the regeneration cavity that $\omega_n = \omega$. The last line contains a fast-oscillating mode that averages out to zero and a slow-oscillating mode. {Finally, we find 
\begin{align}
\label{eq:cavity-Ay}
    & (\partial_t^2 + \gamma \partial_t +  \omega^2) 
     A_{n, y}(t) 
    \\
    & \approx
    \frac{g  B_0 \omega {a_0}}{(k_a - \omega) L}  \;
    [\cos( \omega t- (k_a - \omega) L)  - \cos(\omega t)]
    \cr
    & = 
    g B_0 \omega {a_0} \left[ \frac{2/L}{k_a - \omega }\sin(\frac{k_a-\omega}{2} L) \right]\;
    \sin(\omega t - \frac{k_a - \omega}{2} L) .
    \notag
\end{align}
This will result in the resonant enhancement imposed by $\gamma$ as described in Ref.~\cite{Sikivie:2007qm}.
}

When the magnetic field is rotating, Eq.~\eqref{eq:cavity-const-B} is replaced by two coupled equations
\begin{align}
        (\partial_t^2 + \gamma \partial_t +  \omega_n^2) 
     & A_{n,x}(t) \sin \left (\omega_n z \right )
    \\
    & = 
     g  B_0 \sin(\dot\theta z) \omega {a_0} \cos (k_a z - \omega t),    
\cr
        (\partial_t^2 + \gamma \partial_t +  \omega_n^2) 
     & A_{n,y}(t) \sin \left (\omega_n z \right )
    \cr
    & = 
    g  B_0 \cos(\dot\theta z) \omega {a_0} \cos (k_a z - \omega t),
    \notag
\end{align}
Going through a similar procedure, we reach the equation of motion for $[A_{n,x}(t)~ A_{n,y}(t)~ 0]^T$ as follows:
{\begin{align}
 & (\partial_t^2 + \gamma \partial_t +  \omega^2) 
     A_{n, x}(t) 
    \\
    & = \sum_{\alpha = \pm} \frac{\alpha}{2} g B_0 \omega {a_0} \left[\frac{2/L}{q_\alpha} \sin(\frac{q_\alpha}{2} L)\right] 
    \cos(\omega t - \frac{q_\alpha}{2} L) \notag \\
    \label{eq:cavity-Ay-w-thetadot}
    & (\partial_t^2 + \gamma \partial_t +  \omega^2) 
     A_{n, y}(t) 
    \\
   & = \sum_{\alpha = \pm} \frac{1}{2} g B_0 \omega {a_0} \left[\frac{2/L}{q_\alpha} \sin(\frac{q_\alpha}{2} L)\right] 
    \sin(\omega t - \frac{q_\alpha}{2} L) \,, 
    \notag
\end{align}
where
\begin{align}
 q_\pm =  k_a - \omega \pm \dot \theta \, .
\end{align}
}
In the limit of $\dot\theta$ reducing to zero, it reduces to Eq.~\eqref{eq:cavity-Ay}. 
In the case $\dot\theta \approx \omega - k_a \approx m_a^2/2\omega$, only one of the two source terms is significant. Since $A_{n, x}$ and $A_{n,y}$ differ by $\pi/2$ phase during propagation, we only need to consider one component for the purpose of computing the signal power. Taking $A_{n,y}$ as an example, Eq.~\eqref{eq:cavity-Ay-w-thetadot} is the same as Eq.~\eqref{eq:cavity-Ay}, modulo a factor of two, {after correcting the wave number with $\dot\theta$.} Therefore, we show that the resonance still applies to the LSTW experiments that are enhanced by an optical cavity. 

After this substitution, one can proceed to solve the axion-to-photon conversion probability as in ~\cite{Sikivie:2007qm}.

\section{Additional Experimental Setups}
\label{sec:additional-exp-setup}
As we discussed in the main text, the enhancement of an LSTW experiment can come from a harmonic magnetic field or a rotating magnetic field setup, {both varying either in space or in time}. Aside from designing a specific magnetic profile, we demonstrate similar enhancements that can originate from a modulation of the laser. 

Using the change of reference frame shown in Sec.~\ref{sec:alternative-approach} let us switch to the rotating frame where the magnetic field is constant. The laser that is linearly polarized in the lab frame now has a polarization vector that rotates with $\dot\theta$ frequency. 
\begin{align}
    \vec A(t, z) 
    & = 
    \begin{bmatrix}
        \cos(\dot\theta z) \\
        \sin(\dot\theta z) \\
        0
    \end{bmatrix}
    \mathrm e^{-i(\omega t - k z)}
    \\
    & = 
    \frac{1}{2}
    \begin{bmatrix}
        1 \\
        -i \\
        0
    \end{bmatrix}
    \mathrm e^{-i(\omega t - (k+\dot\theta) z)}  
    +
    \frac{1}{2}
    \begin{bmatrix}
        1 \\
        i \\
        0
    \end{bmatrix}
    \mathrm e^{-i(\omega t - (k-\dot\theta) z)}.
    \notag
\end{align}
To mimic what the rotating magnetic field does, one can directly modify the photon dispersion relation in the lab frame using a material with a refractive index $n\sim 1+\dot\theta/\omega$, which we refer to as the frequency modulation (FM) method. 

{ In addition, modulating the laser amplitude (AM) could have interesting implications for the AMR. Imagine that we modulate the laser as follows
\begin{align}
    \vec A(t, z) 
    & = 
    \begin{bmatrix}
        \cos[\Omega (t-z)] \\
        0 \\
        0
    \end{bmatrix}
    \mathrm e^{-i[\omega (t -  z)]}
    \\
    & = 
    \frac{1}{2}
    \begin{bmatrix}
        1 \\
        0 \\
        0
    \end{bmatrix}
    \mathrm e^{-i[(\omega+\Omega) (t-z)]}
    +
    \frac{1}{2}
    \begin{bmatrix}
        1 \\
        0 \\
        0
    \end{bmatrix}
    \mathrm e^{-i[(\omega-\Omega) (t-z)]}.
    \notag
\end{align}
When there is no laser modulation ($\Omega=0$), axion with mass $m_a = \sqrt{2\omega \dot\theta}$ enjoys a resonant conversion to and from photons. Once laser modulation is included, this will shift the resonance to $m_a = \sqrt{2(\omega\pm \Omega) \dot\theta}$. 
}



Lastly, let us also comment on why simply rotating the laser does not work. One might get the impression that the relative rotation between the magnetic field and the laser machine is the key. As a result, it is intriguing to think of using an optical device to modulate the laser such that the laser machine is effectively rotating while the magnetic field is constant. 
However, if one rotates the laser machine, both the frequency of the photons and the wavelength will be modified. Without introducing the complication due to the ellipticity, let us assume the laser machine has a quarter wave plate so that the photons it emits are circularly polarized. 
Suppose that, at some point in time, the laser emitter outputs a photon whose polarization vector points to the $x$ direction. The wavelength can be determined by the time interval after which another photon pointing to $x$ direction is emitted. Admittedly, rotating the laser machine can lower the frequency of the photons seen by an observer in the lab, $\omega \rightarrow \omega -\dot\theta$, but it will also increase the time interval one has to wait to find another photon that shares the same phase, $k \rightarrow k - \dot\theta$. As a result, the photons' dispersion relation is not modified by a rotating laser machine. 

{
\begin{figure}
    \centering
    \includegraphics[width=.5\textwidth]{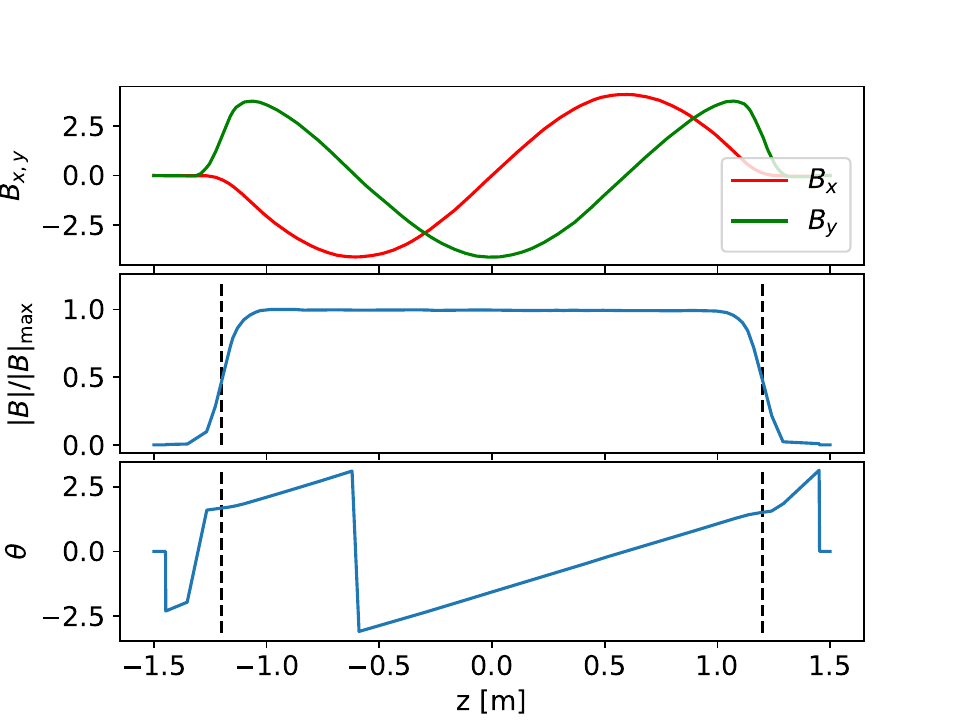}
    \caption{The measured RHIC helical magnet profile~\cite{zotero-21778}.}
    \label{fig:measured-rhic-snake}
\end{figure}
\section{The Fringe Field}
\label{sec:fringe-field}
One might get worried about the irregularity of the magnetic field at the edge of the helical magnets. In this section, we verify that its impact on the resonance is minimal.

The fringe field of the RHIC snake magnets extends about 10~cm beyond the ``regular'' sinusoidal component at both ends \cite{zotero-21781,753208,zotero-21778,zotero-21777,Anerella:2003se,anerella1999superconducting,Alekseev:2003sk}.
Heuristically, this configuration has a constant helical frequency $\dot \theta$ in most of the optical cavity while the resonance is lost in the small fraction of the total propagation length due to the edge effect of the magnets. Therefore, its impact on the total conversion rate between axion and photons is completely negligible.

To back up with this argument, we perform two numerical tests. First of all, add a 10~cm gap between any two magnets (2.4~m each). In the gap, we set the magnet field to zero to maximize the ``disruption'' of the resonance. The result in Fig.~\ref{fig:fringe} shows that its impact on the reach of $g_{a\gamma}$ is negligible. 

Next, we take the actual magnetic field profile from \cite{zotero-21778}. The profile is shown in Fig.~\ref{fig:measured-rhic-snake}.
Since the laser beam has a beam radius at the order of $5\mathrm{mm}$~\cite{Bahre:2013ywa}, the sextupole and higher multipole terms are negligible. We again find that the fringe field impact on the AMR is negligible. The result is included in Fig.~\ref{fig:fringe}.
\begin{figure}
    \centering
    \includegraphics[width=.45\textwidth]{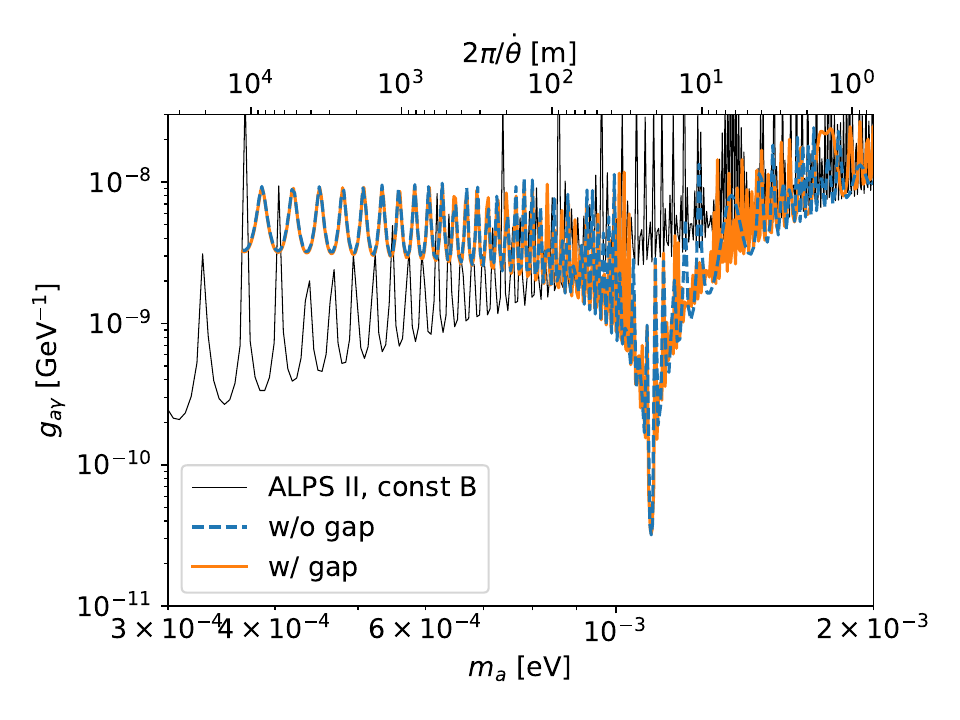}
    \includegraphics[width=.45\textwidth]{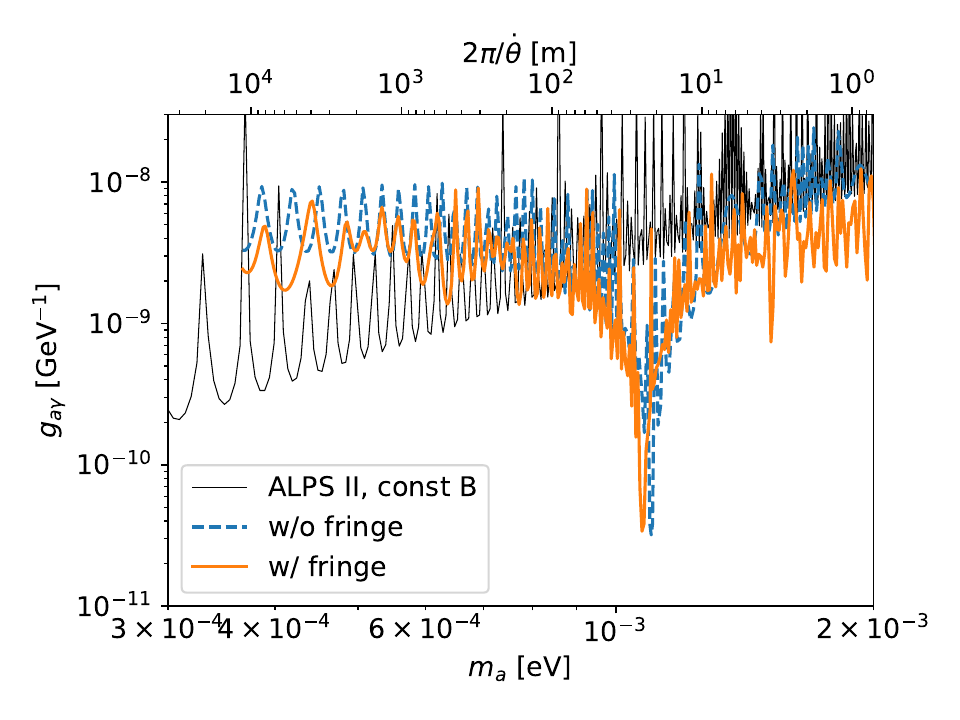}    
    \caption{We perform two numerical tests to verify that the fringe field of the RHIC helical magnets has minimal impact on the resonance. \textbf{Above}: we manually add a 10~cm gap where magnetic field is zero. \textbf{Below}: we use the measured RHIC magnetic profile and take into account the fringe field at both ends assuming a 10~cm gap between any two magnet modules.}
    \label{fig:fringe}
\end{figure}

Lastly, we comment that the fringe field of the RHIC helical snake has a sizable longitudinal components around the axis~\cite{zotero-23753}. However, this component does not contribute to the axion-photon conversion in vacuum. Therefore, it has a much smaller impact for LSTW experiments compared to that when used to control polarized beams.

\section{Integer Resonance Modes in the Discrete Helical Array}
\label{sec:high-modes}
When adopting the discrete helical array, the effective helical frequency is determined by 
$\dot \theta = \Delta \theta/ \Delta L$, where $\Delta L$ represents the size of the dipole module.
Since $\Delta \theta$ is defined on $[0, 2\pi)$, it coincides with $\Delta \theta + 2\pi \, n$ with $n$ being any integer. In other words, the discrete helical array not only contains the helical frequency $\Delta \theta/ \Delta L$, but also includes higher modes:
\begin{align}
\label{eq:high-modes}
    {\dot \theta_n}
    & = 
    \frac{\Delta \theta}{\Delta L} \pm \frac{2\pi}{\Delta L} n,
\end{align}
where $n = 0, 1, ...$. 
This interesting feature enables us to reach even higher axion masses. Using this formula, we can easily verify the origin of the higher resonance modes as shown in Fig.~\ref{fig:high-modes}.
\begin{figure}
    \centering
    \includegraphics[width=.5\textwidth]{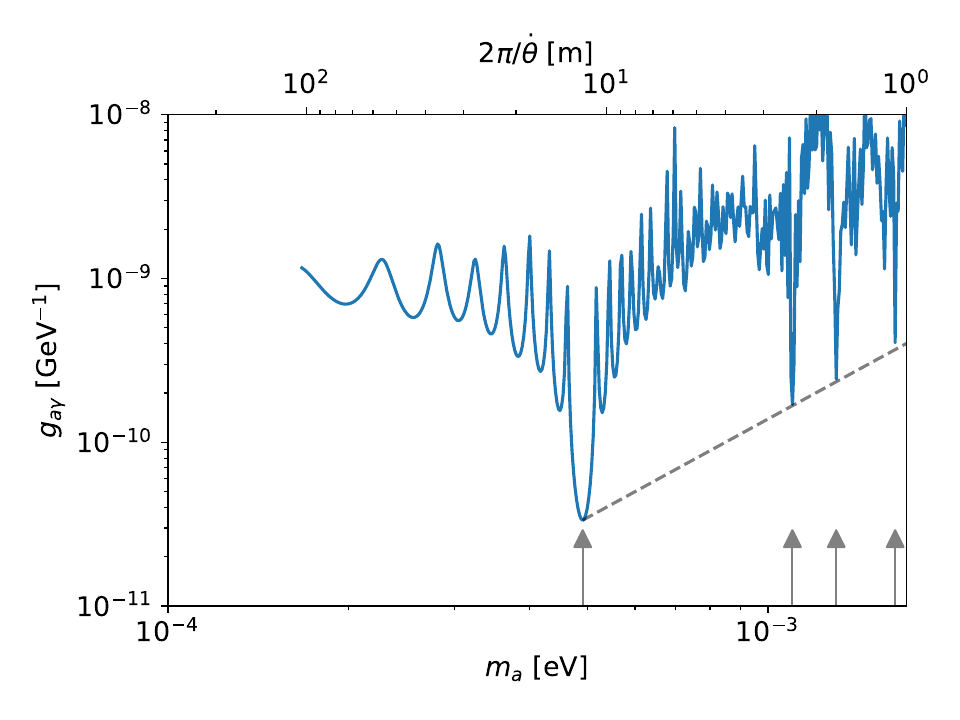}
    \caption{The integer resonance modes present in the discrete helical array. The grey arrows are put to the locations according to Eq.~\ref{eq:high-modes}. They match with the resonance troughs from the numerical computation. The gray dashed curve is the predicted depth as shown in Eq.~\ref{eq:depth-estimate}.}
    \label{fig:high-modes}
\end{figure}

The height of the higher resonance modes ($n\neq 1$) can also be estimated analytically. 
The magnetic field can be transformed into the frequency domain. However, due to the ambiguity of its frequency, we can expand it into different Fourier bases. For example, $n=0$ contains the following Fourier modes:
\begin{align}
    \mathrm{e}^{-i \dot \theta_0 z}, \mathrm{e}^{-i 2\dot \theta_0 z}, \mathrm{e}^{-i 3\dot \theta_0 z}, \cdots
\end{align}
For $n=1$, it contains the following modes
\begin{align}
    \mathrm{e}^{-i \dot \theta_1 z}, \mathrm{e}^{-i 2\dot \theta_1 z}, \mathrm{e}^{-i 3\dot \theta_1 z}, \cdots
\end{align}
In what follows, we will take the lowest mode in each basis because all the modes with integer multiples are highly integrated out:
\begin{align}
    B(z)
    & = \sum_{n, m} b_{n,m} \mathrm{e}^{-i \, m\, \dot{\theta}_n z}
     \simeq \sum_{n} b_{n} \mathrm{e}^{-i\dot{\theta}_n z},
\end{align}
where $b_{n} \equiv b_{n,1}$, and the real and imaginary parts of $B(z)$ correspond to the $x$ and $y$ components of the magnetic fields.
We stress that the summation is not over a single Fourier basis but summing over the lowest mode of each Fourier basis. 

Meanwhile, notice that the frequency $\dot\theta_n$ is only an average. In each dipole magnet the magnetic field is approximately constant. Let us check the overlap between a constant magnetic field and each lowest Fourier mode. 
\begin{align}
\label{eq:depth-estimate}
    b_{n}
    & = \frac{1}{L}\int_0^{L} dz \; B(z) \; \mathrm{e}^{i \dot{\theta}_n z}
    \cr
    & = 
    \frac{1}{L}
    \sum^{N-1}_{k=0}
    \int _{k\Delta L}^{(k+1) \Delta L}
    dz \;
    |B| \, \mathrm{e}^{-i \,k \Delta \theta}
    \mathrm{e}^{i\dot \theta_n z}    
    \cr
    & = 
    \frac{1}{L}
    \sum^{N-1}_{k=0}
    \int _0^{\Delta L}
    dz'\;
    |B|\, \mathrm{e}^{-i\, k \Delta \theta}
    \mathrm{e}^{i\dot \theta_n(z' + k\Delta L)}
    \cr
    & = 
    \frac{1}{\Delta L}
    \int_0^{\Delta L}
    dz' \; |B|\, 
    \mathrm{e}^{i \dot \theta_n z'}
    \cr
    & =
    \frac{B}{\dot{\theta}_n} 
    \left (\frac{\mathrm e^{i \Delta \theta} - 1 }
    {i\Delta L}
    \right )
    \propto
    \frac{B}{\dot \theta_n}
\end{align}
where $B$ is the magnetic field amplitude of the discrete helical array, and the integration is performed over one period of that array. This tells us that the overlap between the constant magnetic field of each dipole and a $\dot \theta_n$ mode gets smaller as $n$ gets larger. Put differently, the integral contains $n$ sinusoidal periods, which gives no contribution to $b_n$, and a small remaining length gives contributions the same way as the $\dot\theta_0$ mode but with a reduced length. This leads to $b_n / b_0 = \dot{\theta}_0/\dot{\theta}_n$.
} Therefore, the magnetic power decreases for higher modes. This heuristic argument is verified by the numerical solution in Fig.~\ref{fig:high-modes} as shown by the gray dashed curve.


\section{Comparison with Previous Works}
\label{sec:comparison}

We highlight a few key differences between our work and previous works where the variation of a magnetic field is suggested.

{In \cite{Sikivie:1983ip}, a sinusoidal modulation of the magnetic profile was first mentioned as a way to enhance the solar axion search. We note that while this can lead to great enhancement in the axion-to-photon conversion, LSTW experiment enjoys far better resonant improvement due to the flux being mono-energetic.}
In \cite{VanBibber:1987rq}, a segmented magnetic field configuration of alternating polarity, the so-called the \textit{Wiggler}, was suggested to scan the heavier axion masses in the original paper. The paper cast the separated magnetic dipoles in terms of a form factor and sketched out this configuration's potential in extending the reach of axion searches. In \cite{Arias:2010bha,Arias:2010bh} more systematic studies were performed for the alternating dipole design. In addition, a magnet array with gaps in-between was studied. 
We note that the AMR studied in this Letter is more generic. It not only explains the Wiggler/magnet-with-gap setup using approaches different from the original work, but also points to the optimal setup that can create the resonance, namely a magnetic field with a {constant rotation in its orientation perpendicular} to the photon propagation direction. 
Besides, the AMR applies to a rotating/oscillating magnetic field temporally/spatially, as well as more generic setups such as the laser modulations as we show. We also discuss how the magnetic field induced enhancement is related to the usual method, where the cavity is filled with gas/plasma, in the context of the AMR formalism. 
{

{
We note in the main text that the AMR can be applied to a discrete setup dubbed ``discrete helical array''. In this relaxed helical setup, the Wiggler proposal is a special case by setting the relative axial angle $\Delta \theta = \pi$. It leads to a considerable improvement over the Wiggler setup as shown in Fig.~\ref{fig:discrete-helical-vs-Wiggler}.
This example serves as an example of the great potential of AMR and its experimental relevance.
Here, we use HERA dipole magnets to induce the helical profile and the AMR. As a result, it is no longer a competing design to the Wiggler proposal. We hope that it will motivate the experimental community to reconsider the Wiggler proposal with a potential future upgrade to allow different magnets' axial orientation to be adjustable.
We comment in passing that the discrete helical array can bypass the need of inverting the electric current in each superconducting magnet, which is the main challenge to implementing it at ALPS II due to the use of diodes. 
}

\begin{figure}
    \centering
    \includegraphics[width=.5\textwidth]{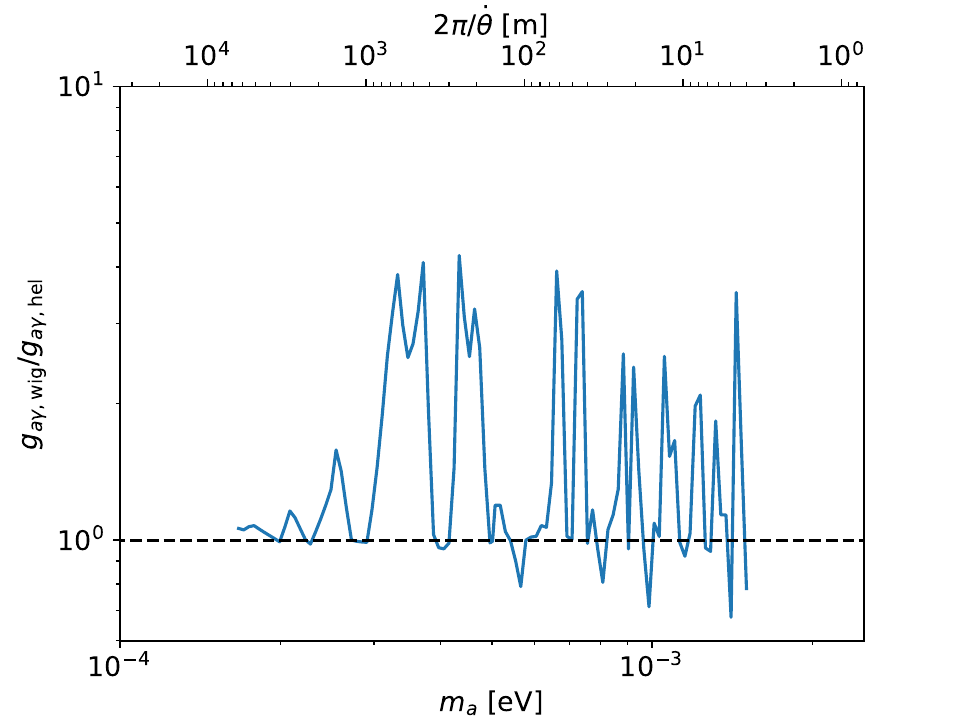}
    \caption{Comparison of the experimental reach of $g_{a\gamma}$ using the Wiggler method and the discrete helical array.}
    \label{fig:discrete-helical-vs-Wiggler}
\end{figure}

In \cite{Jaeckel:2007gk}, an idea using phase-shift plates to extend the experimental reach in the LSTW setup is also suggested instead of varying magnetic fields. This modifies the photon fields directly. We note that the phase modulation at the wave plate serves as an ``external kick" that introduces extra phase between the photon field and the external magnetic field. 
This, in turn, can be understood using the resonance within the AMR framework. 

In the experiment PVLAS~\cite{zotero-21715} a rotating magnetic field is used. However, the rotation of the magnetic field in ~\cite{zotero-21715} is not to create a resonance with the axion-photon momentum transfer but to reduce measurement noise. 

The work in \cite{Zarei:2019sva,Sharifian:2021vsg} discusses how a varying magnetic profile affects the vacuum birefringence searches, \textit{i.e.} PVLAS. First of all, our approach focuses on the luminosity of the photon-axion conversion, hence is relevant to LSTW and solar experiments, while \cite{Zarei:2019sva,Sharifian:2021vsg} studies the axion-induced phase shift in the context of PVLAS. 

In addition, \cite{Zarei:2019sva,Sharifian:2021vsg} studies the pulsed magnetic field setup while we discussed both the pulsed (harmonic) setup and the helical profile, which turns out to be even more effective than the harmonic profile. 

While the authors of \cite{Zarei:2019sva,Sharifian:2021vsg} use second quantized fields to compute the effect, we establish that a simple way to understand that such a resonance can be achieved through the pure classical approach. In addition, we verify it using different pictures including the avoided level crossing~(Eq.~\eqref{eq:diagonalize-12}), the parametric resonance~(Eq.~\eqref{eq:parametric-resonance-helical}), and the preferred non-inertial reference frame~(Fig.~\ref{fig:inertial-frame-vs-non-inertial-frame}). We provide a simple way to understand the resonance in the harmonic magnetic profile by decomposing it into two helical profiles in Eq.~\eqref{eq:harmonic-decompose-helical}. 

We also demonstrate the technological feasibility of a helical magnetic profile with the specification of the RHIC helical magnets.

{

Similar formalism was also adopted in the analysis of neutrino oscillations. See, for example, Refs.~\cite{Aneziris:1990my,Smirnov:1991ia,Akhmedov:1993ta,Akhmedov:1993sh,zotero-1712} and references therein.
}
\bibliography{bib}
\bibliographystyle{utphys}
\end{document}